\documentclass[num-refs]{wiley-article}

\usepackage{siunitx}
\usepackage{pbox}
\usepackage{subcaption}

\papertype{Original Article}
\paperfield{Journal Section}

\title{A Pilot Study For Fragment Identification Using 2D NMR and Deep Learning}

\author[1]{Stefan Kuhn}
\author[ ]{Eda Tumer}
\author[1]{Simon Colreavy-Donnelly}
\author[2]{Ricardo Moreira Borges}

\affil[1]{School of Computer Science and Informatics, De Montfort University, Leicester, UK}
\affil[2]{Instituto de Pesquisas de Produtos Naturais Walter Mors, Universidade Federal do Rio de Janeiro, Brazil}

\corraddress{Stefan Kuhn, School of Computer Science and Informatics, De Montfort University, The
Gateway, Leicester LE1 9BH, UK}
\corremail{stefan.kuhn@dmu.ac.uk}

\fundinginfo{De Montfort University, VC2020 L SL 2020}

\runningauthor{Stefan Kuhn et al.}

\begin{document}

\begin{frontmatter}
\maketitle

\begin{abstract}
This paper presents a method to identify substructures in NMR spectra of mixtures, specifically 2D spectra, using a bespoke image-based Convolutional Neural Network application. This is done using HSQC and HMBC spectra separately and in combination. The application can reliably detect substructures in pure compounds, using a simple network. It can work for mixtures when trained on pure compounds only. HMBC data and the combination of HMBC and HSQC show better results than HSQC alone. 



\end{abstract}
\end{frontmatter}

\section{Introduction}
\label{sec:introduction}

Nuclear Magnetic Resonance Spectrometry (NMR) is an established technique in structure elucidation. Working with pure compounds, it is now possible to determine structures from NMR data reliably and economically, either using knowledge and experience or relying on computer software \cite{ELYASHBERG201588}. The situation is different when it comes to mixtures, though. When a complex mixture is measured, determining the components of that mixture is in many cases difficult or impossible \cite{LEGGETT2019407}. A separation of the compounds can be attempted, but this is time-consuming, expensive, and potentially destroys or modify some of the compounds. Therefore, a direct structure elucidation from the mixture would be desirable. Potentially, separating the peaks in the spectrum by compound would be an important part of the solution, reducing the structure elucidation of those compounds to a known problem. Unfortunately, to do so is a difficult. On the other hand, if the compounds is known, assigning the peaks to them might be relatively easy. Of course, this is a circular situation. Breaking the circle could potentially be possible by identifying substructures and functional groups. A relatively small number of those should be enough to identify a number of peaks belonging together in any mixture of typical organic compounds. Once those fragments have been identified, it should be easier to tackle the task of separating the compounds for the structure elucidation of the unknowns.

As the first step for this, an algorithm is used to classify 2D NMR spectra and therefore the underlying compounds by substructures they contain. For this, a deep learning approach (DL from here) is used, which have shown their power in a number of applications since they were developed. In this study, classification only of spectra by occurrence or non-occurrence of fragments is considered. A next step will be identifying the peaks belonging to those fragments. From there, work can continue on the task of mixture segmentation outlined above. The specific DL method used is a Convolutional Neural Network (CNN), trained on image data. The main work in the area so far is SMART-NMR \cite{pmid29079836, doi:10.1021/jacs.9b13786}, which uses HSQC spectra and a CNN approach to cluster structures and find the most similar existing structure to a spectrum of an unknown compound.
In contrast to SMART-NMR, this project explores the use of HSQC and HMBC for structures containing a certain fragment. 
Thus, this DL image-based method is intended to focus on identifying single predefined structures in this first attempt. To access the value of this approach mixtures of compounds containing candidate substructures are evaluated.



\section{Background}

\subsection{Metabolomics and spectroscopy}

The role that metabolomics play together with the other 'omics' technologies in the understanding of complex biological systems is broadly accepted. Many Life Science projects have implemented untargeted metabolomics within their pipeline to help them characterize phenotypes for different applications \cite{Lindsay2020, Penaloza2020, Feussner2019, Greer2011, Morais2020, Rutz2019, Borcherte00843, Edison2020Unique}. However, one of the major bottlenecks to be overcome in this approach is compound identification, since important features with no identification lead to no meaningful understanding of a biological system \cite{Garcia2020, Sindelar2020}. Even though most of the metabolomics applications have been done using mass spectrometry (MS)-based platforms, nuclear magnetic resonance (NMR) has complimentary advantages that are worth noting \cite{LEGGETT2019407, Edison2020Unique}, such as its quantitative response and reproducibility  \cite{MARKLEY201734}. All in all, the Metabolomics Standards Initiative (MSI) propose the use of confidence levels for compound annotation where the only accepted full identification (or level 1) is achieved with a commercial authentic sample \cite{Sumner2007}, whereas level 2 represents putatively annotated compounds by spectral similarity using MS or NMR.

Annotation tools for MS data have developed significantly in the last years \cite{Nothias2020, Rogers2019, Ludwig2018, Silva2018}, but users are limited to the intrinsic limitations imposed by e. g. isobaric structures. The identification confidence can be increased by adding orthogonal information such as the fragmentation pattern and retention time when a chromatography separation is used \cite{Bonini2020,Naylor2020}. Notably, the use of NMR data in this scheme is still under-explored in MS-based studies \cite{Kuhn2019}. The MS side is not the focus of this discussion, instead, there is an ongoing effort to develop methods for NMR-based compound identification within mixtures to complement the already well-advanced MS-based tool. The idea is to suggest the acquisition of a set of 2D NMR spectra of a studied sample to enable another factor of confidence for compound identification. Through that, to create a virtuous circle to increment structural elucidation propagation of close related MS/MS spectra e.g. from a Molecular Networking analysis \cite{pmid33222074}. 

\subsection{Deep learning, image processing, and NMR}

Digital image processing (see \cite{gonzalez2018digital} for an overview) is an established branch of computer science, which has taken a leap forward over the recent years with the introduction of deep learning (see \cite{8744516} for a survey on deep learning and image analysis). The tasks performed by such systems include image classification (images are classified according to their content), image segmentation (dividing images into constituent parts), or object detection (assigning labels to parts of the image). Clearly, these and other tasks are interlinked. Typical application areas include metrology, medical diagnosis, remote sensing, autonomous vehicles, and many more. NMR spectra are, in the first instance, signals in time domain. Those get transformed into frequency domain spectra and processed by a number of methods, e. g. baseline correction, to produce the spectrum which is normally presented, in a visual form, as an NMR spectrum. An analysis of NMR data could, therefore, apply at many different levels. The hypothesis is that using image representations, and treating them with the same techniques used in other image processing applications, can yield insights into the chemical properties of the substances behind the spectra. Specifically, we use CNNs as the dominant technique in deep learning-based image analysis, for finding chemical information in 2D NMR spectra. In this paper, classification techniques are used to classify spectra by fragments contained in the structure they represent. As outlined in Section~\ref{sec:introduction}, identification of the peaks will be a next step, which would be an image segmentation task in image processing terminology.

Deep learning techniques, as well as traditional machine learning techniques, have been used in conjunction with NMR for other applications, for an overview see \cite{doi:10.1002/mrc.4989}. Some of these areas are not directly related to the exercise (like prediction of 1D shifts, recently done in \cite{pmid31388784}), but the area of peak picking (e. g. \cite{10.1093/bioinformatics/bty134} employs a CNN for peak classification) uses imaging-related technologies as well. SMART-NMR \cite{pmid29079836, doi:10.1021/jacs.9b13786} uses CNNs for similarity detection (see Section~\ref{sec:discussion}). Computer-aided structure elucidation systems (e. g. SENECA \cite{seneca}) are superficially similar to the task, but they work on pure compounds and normally use optimization techniques, which is not feasible for substructures and complex mixtures. Finally, in biomedical analysis \cite{LINDON20011} and metabolomics \cite{pmid31628551} neural networks are used for pattern matching and similarity scoring, but not using the image representations.

\section{Methods}
\begin{figure}
\begin{subfigure}{.3\textwidth}
  \centering
  \includegraphics[width=.8\linewidth]{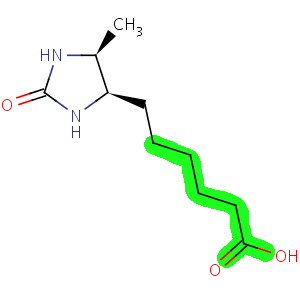}
  \caption{A fatty acid (bmse000314)}
  \label{fig:structure1}
\end{subfigure}%
\begin{subfigure}{.3\textwidth}
  \centering
  \includegraphics[width=.8\linewidth]{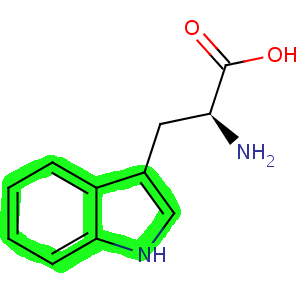}
  \caption{An indole (bmse000050)}
  \label{fig:structure2}
\end{subfigure}d
\begin{subfigure}{.3\textwidth}
  \centering
  \includegraphics[width=.8\linewidth]{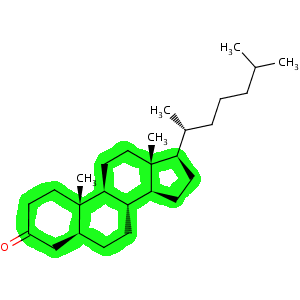}
  \caption{A steroid (bmse000489)}
  \label{fig:structure3}
\end{subfigure}
\caption{Example for the three categories of structures, with the characteristic substructure highlighted.}
\label{fig:structures}
\end{figure}

The main source of data for this work is the Biological Magnetic Resonance Data Bank \cite{bmrb}. This project provides spectral data of 1732 small molecules in \url{http://www.bmrb.wisc.edu/ftp/pub/bmrb/metabolomics/entry_directories/}. We have selected from those (using the given .mol files) molecules containing certain substructures (Figure~\ref{fig:structures}). For the fatty acids, the hydrogen atoms were included in the substructure search, making sure there are no side chains (except on the terminal carbon atom of the fragment). For the other substructures, no hydrogen atoms were included, allowing side chains in any position. In addition, selected molecules from nmrshiftdb2 were chosen, using the same substructures, \cite{nmrshiftdb2} which have full spectra. From this is derived 29 HMBC and 33 HSQC spectra (not all structures have both) for fatty acids, 25 HMBC and HSQC spectra for steroids, and 28 HMBC and 32 HSQC spectra (not all structures have both) for indole.

For all those structures, the image representations of the HMBC and HSQC spectra are retrieved. This means that the representation depends on the settings of the export process. This is an acceptable situation, and one aim of this paper is to show that this can be overcome by a CNN. Some preprocessing steps have been performed. The steps are:

\begin{itemize}
    \item Conversion to jpeg
    \item Conversion to greyscale if colours were used
    \item Removal of scales and grids (there would be danger of those being picked up by the network)
    \item Scaling to 600x410 pixels
    \item The image area was extended (or in some cases cropped) to cover the area of 0-10/0-200 ppm. This was only done approximately, so it is not to have exact matches.
\end{itemize}

\begin{figure}
\begin{subfigure}{.33\textwidth}
  \centering
  \frame{\includegraphics[width=.9\linewidth]{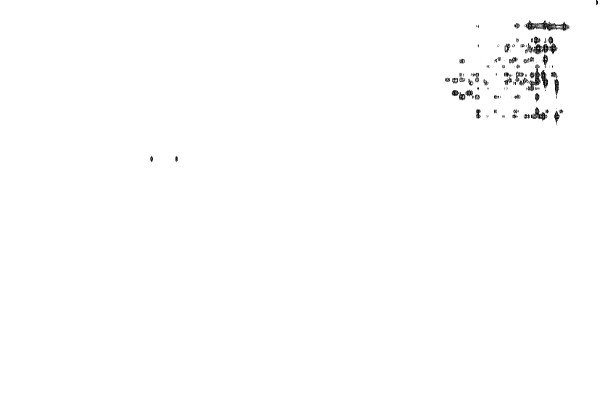}}
  \caption{}
\end{subfigure}%
\begin{subfigure}{.33\textwidth}
  \centering
  \frame{\includegraphics[width=.9\linewidth]{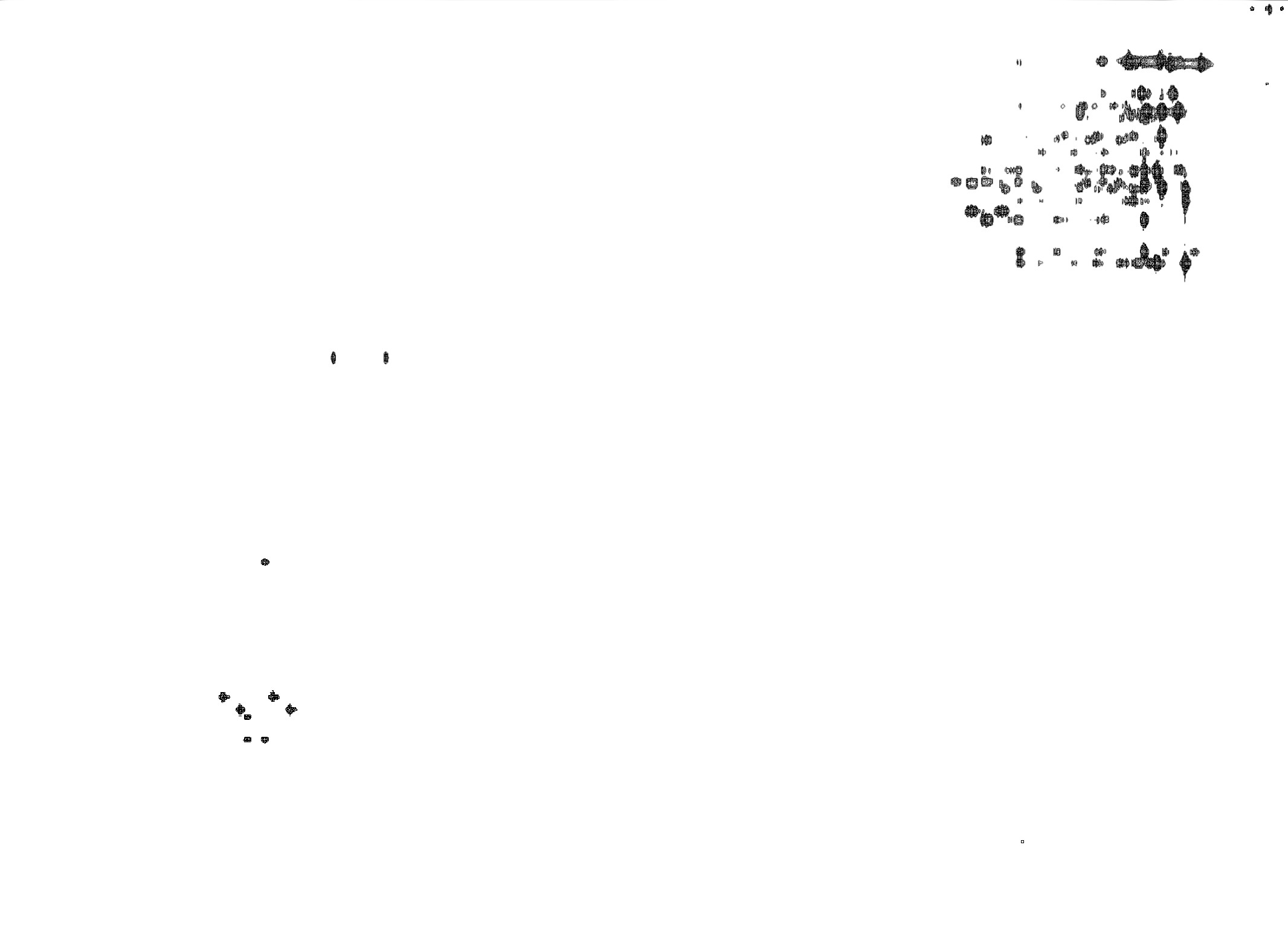}}
  \caption{}
\end{subfigure}
\begin{subfigure}{.33\textwidth}
  \centering
  \frame{\includegraphics[width=.9\linewidth]{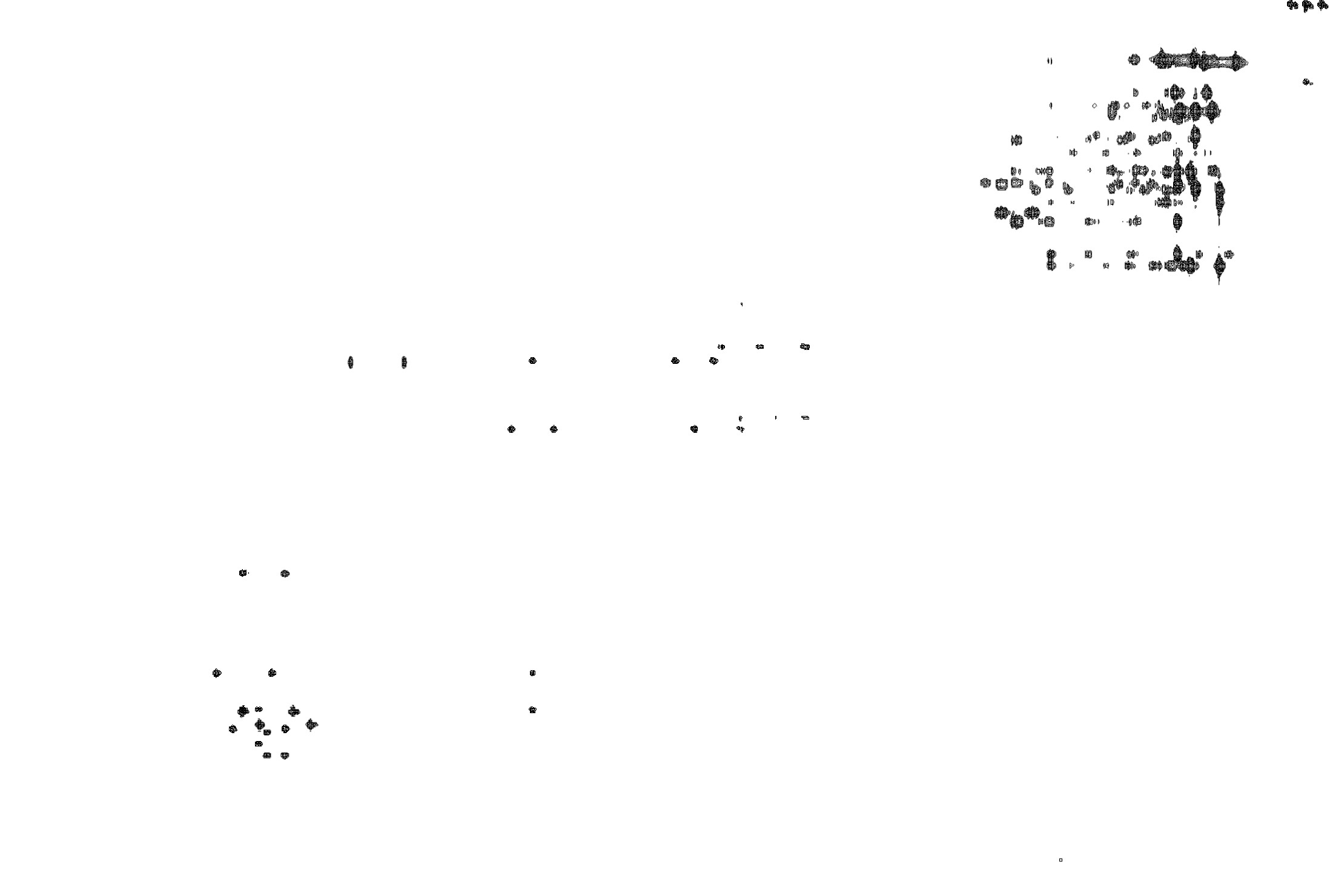}}
  \caption{}
\end{subfigure}

\begin{subfigure}{.33\textwidth}
  \centering
  \frame{\includegraphics[width=.9\linewidth]{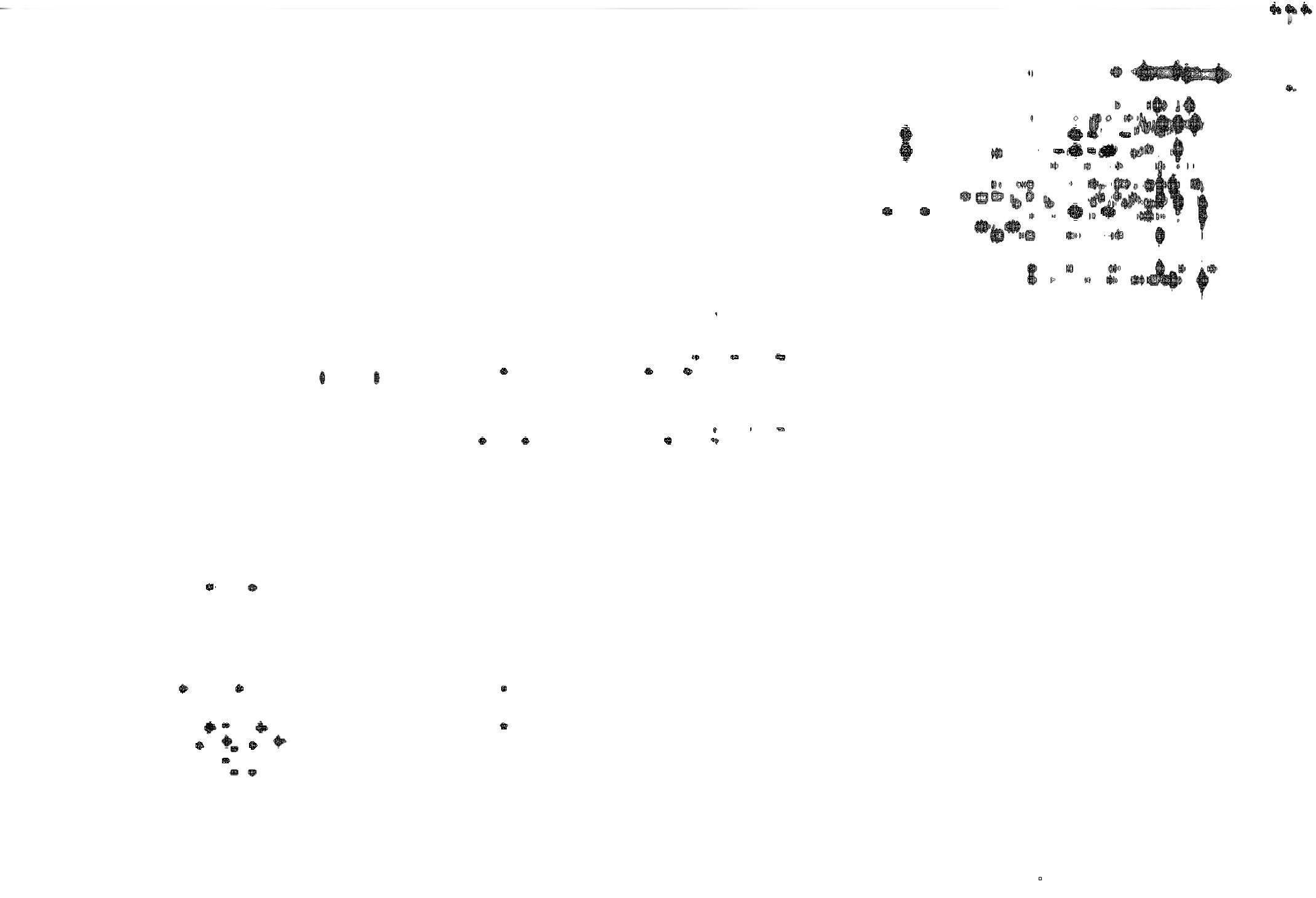}}
  \caption{}
\end{subfigure}%
\begin{subfigure}{.33\textwidth}
  \centering
  \frame{\includegraphics[width=.9\linewidth]{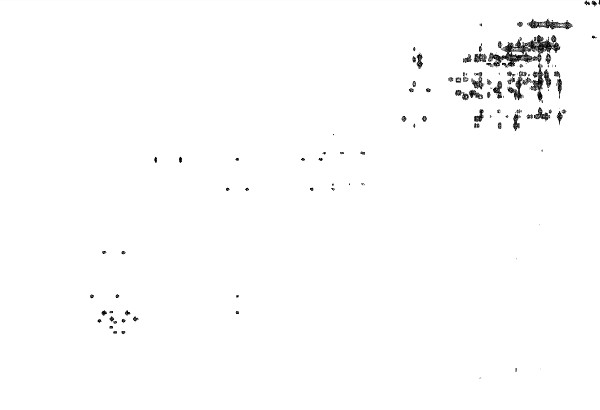}}
  \caption{}
\end{subfigure}
\begin{subfigure}{.33\textwidth}
  \centering
  \frame{\includegraphics[width=.9\linewidth]{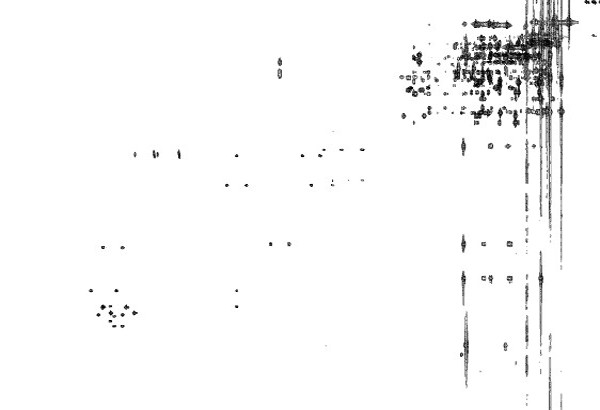}}
  \caption{}
\end{subfigure}
\caption{The HMBC spectrum of bmse000489 (a), overlayed with an increasing number of other HMBC spectra.}
\label{fig:mixtures}
\end{figure}

A number of 'artificial mixtures' were also created for each class of compounds. For those, the spectral images were overlayed from the compounds bmse000643 (fatty acid; elaidic acid), bmse000364 (indole; 5-hydroxyindole-3-acetic Acid), and bmse000489 (steroid; 5-alpha-cholestan-3-one) with a succession of other spectra. The compounds used for overlaying have been randomly chosen and are bmse000060 (adenine), bmse000061 (adenosine), bmse000072 (cadaverine), bmse000491 (alpha-pinene-oxide), and bmse000544 (cholesteryl palmitate), in this order to create 5 mixtures for each substructure case. The scales of each image were used to ensure exact fit. The images were treated with the steps outlined above. This was done separately for HSQC and HMBC spectra. For instance, figure~\ref{fig:mixtures} shows the resulting spectra for the HMBC spectrum of bmse000489 of each complexity increment. Altogether, this gives 15 mixtures, which are used for evaluating the ability of the neural network to deal with mixtures. 

All generated images are available in the github repository (see Supporting Information).

The proposed networks are developed with Python 3 using Keras and Tensorflow libraries \cite{keras, tensorflow}. As part of the github repository (see Supporting Information), is provided containing a list of required packages and a README file for the installation. The network which is used for the classification of a single spectral image by processing HMBC or HSQC spectra consists of 7 layers, 2 of which are input and output layers (Figure~\ref{fig:structure1}). This network has been trained with HMBC and HSQC separately. Its accuracy rate can be found in Table~\ref{table:pure_and_mixtures}.
Initially, the custom model consisted of convolutional layers with depths 32-64-64  which were then connected to a dense layer with a depth of 64. The problem of this model was the randomness of its results. It often tended to be stuck at a low accuracy such as 0.305 at an early stage of the training nonetheless the epochs advanced. The inconsistency and instability of this model have improved when the layer with a depth of 128 was introduced right before the dense layer.

\begin{figure}
\begin{subfigure}{.5\textwidth}
  \centering
  \includegraphics[width=.9\linewidth]{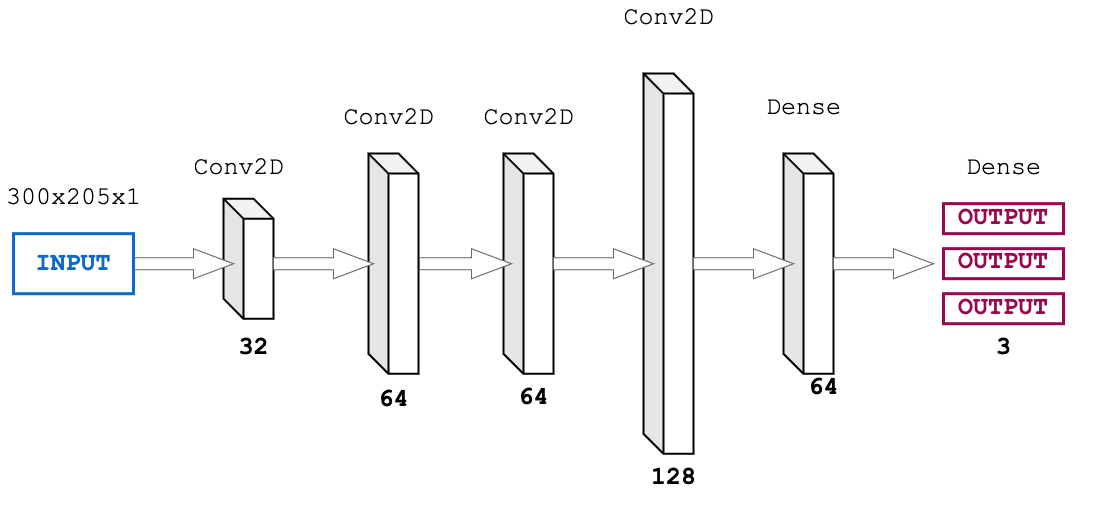}
  \caption{}
  \label{fig:architecture1}
\end{subfigure}%
\begin{subfigure}{.5\textwidth}
  \centering
  \includegraphics[width=.9\linewidth]{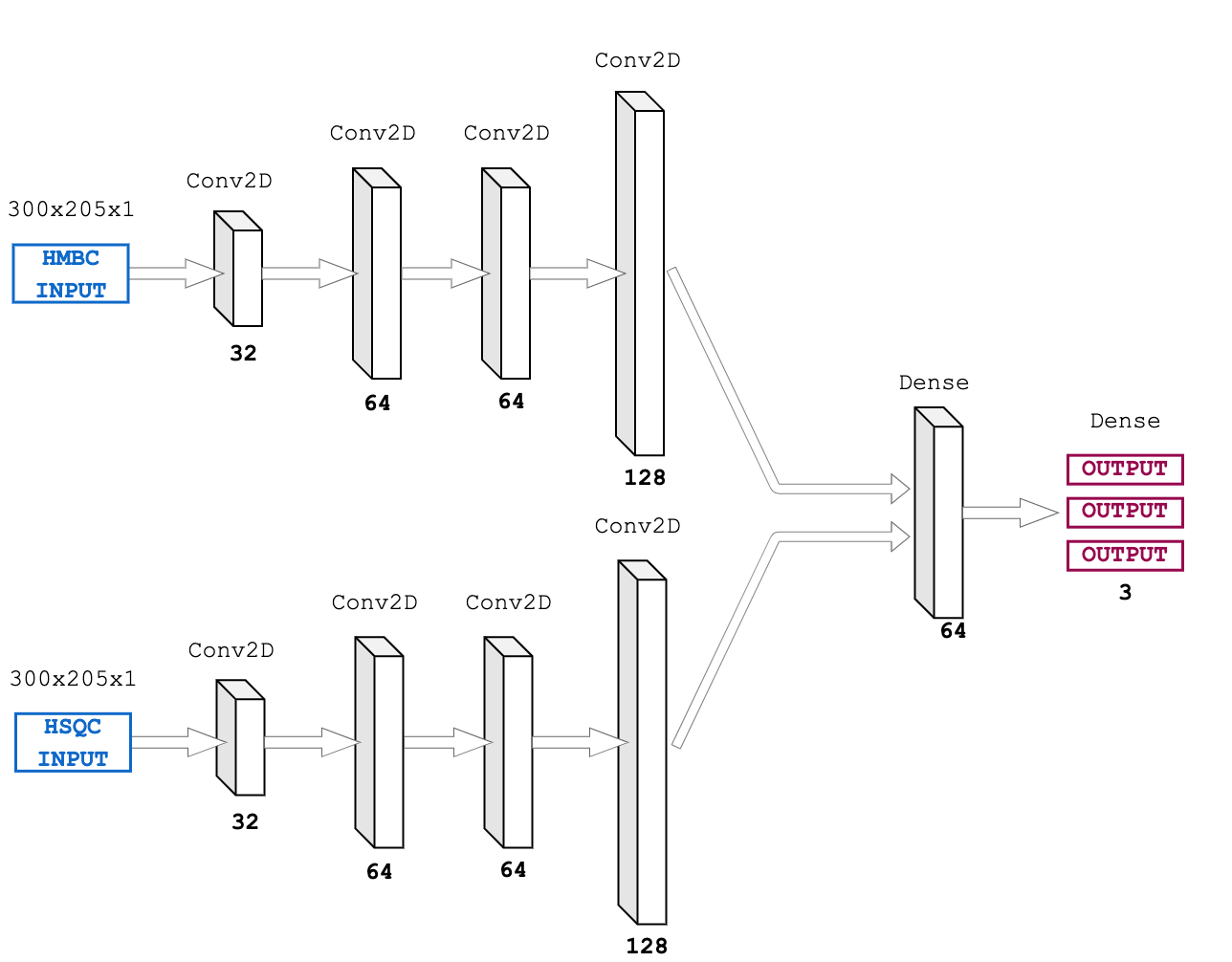}
  \caption{}
  \label{fig:architecture2}
\end{subfigure}
\caption{Architecture of (a) the single spectrum convolutional neural network and (b) the convolutional neural network for combined HMBC and HSQC spectra.}
\label{fig:architecture}
\end{figure}

After various experiments with different architectures and hyper-parameters, It was concluded concluded that the architecture presented in Figure~\ref{fig:architecture1} has been the attempt that responded well to NMR spectra for both HMBC and HSQC images. Therefore, the same structure is also used in the combined neural network (Figure~\ref{fig:architecture2}).
The other attempts include investigating an architecture which is proposed for peak picking \cite{10.1093/bioinformatics/bty134}, a classification model published by Tensorflow (depths of layers 16,32,64,128,3) \cite{tfimgclassonline}, and a variant of the Tensorflow model with a smaller dense layer (depths of layers 16, 32, 64, 64, 3). 

The proposed combined CNN architecture consists of 2 initial branches which are then connected to a dense layer with a depth of 64 and followed by an output layer which is a dense layer with a depth of 3. The branches are composed of 4 Conv2D layers with a kernel size of 3x3 and max pooling with a kernel size of 2x2 is applied after each of these layers. The output of the architecture is then reduced (via Softmax layer), to a confidence estimation of essentially 0 or 1, as is the accepted way in image analysis. It should be noted that the network is never trained with spectra of structures containing none of the substructures, and also not with mixtures of them. This, together with the Softmax architecture, restricts the network to a classification to contain one of the substructures. Both networks were optimised by manually sampling the most important parameters of the network (number of layers, type of layers, filter size and stride of convolutional layers, size of max pooling layers). The result is the networks given. Due to the scarcity of data, it was decided to avoid using more complex architectures such as some of the winning competitors in the latest ImageNet competitions, e. g. the current leader \cite{foret2020sharpnessaware}. At this stage, it is necessary to show that the application of CNNs to this problem works in principle. K-fold cross-validation is used, specifically stratified k-folds, to evaluate the model. This is used to ensure that pure substances can be classified, as the first step to mixture analysis. For the mixtures, the artificial mixtures mentioned before are used rather than a full cross-validation. This is because producing the mixtures for each compound would be very difficult. The training for classifying the mixtures was done using all pure compounds except those three contained in the mixtures.

\section{Results}

It is shown that a simple CNN, a type of network which is commonly used for image analysis, is able to classify 2D NMR spectra of pure compounds according to substructures present. This has been shown by using cross-validation on the full set of data. It is established the number of folds to use by trying 2, 3, 5, and 10 fold cross-validation. The results for this are shown in Table~\ref{table:kfolds}, using the combined HSQC and HMBC network. This shows considerably satisfactory results already with 2 folds, but slightly better results with 10 folds. Therefore 10-fold cross-validation is done for use of HMBC only, HSQC only, and the combined network. The results of this are shown in the first line of Table~\ref{table:pure_and_mixtures}. HMBC only and the combined network give an accuracy of over 90\%, whereas the HSQC only network achieves a lower accuracy of approximately 83\%.

\begin{table}[h!]
\begin{center}
 \begin{tabular}{||c c||} 
 \hline
  Folds & Accuracy \\ 
 \hline\hline
 2 & 87.92\% \\ 
 \hline
 3 & 90.01\% \\ 
 \hline
 5 & 86.63\% \\ 
 \hline
 10 & 90.84\% \\ 
 \hline
\end{tabular}
\caption{Accuracy of predicting compound classes for pure compounds, using cross-validation with different number of folds. All numbers are for combined HSQC and HMBC.}
\label{table:kfolds}
\end{center}
\end{table}

Based on this, it was also examined if a network, which was trained on pure compounds only, can distinguish substructures in mixtures. The mixtures contained at least one compound that the model was trained with. The results for the 15 mixtures for each type of network are shown in the second line of Table~\ref{table:pure_and_mixtures}. As for the pure compounds, HSQC alone performs worst. HMBC alone is better than the combined network, but due to the low number of samples these numbers cannot be used for a definitive judgement. It should be noted that the underlying pure compounds were correctly detected in all cases. Due to the low number of samples, these results should be treated with caution. Yet, this is an indication that learning from pure samples can be successfully applied to mixtures.

Table~\ref{table:performance_mixtures} gives a more detailed view of the results for mixtures. It shows that the errors are partly consistent, so for example using HMBC only the steroid is not found very well, but it is consistently considered an indole. This seems to indicate that the results are not purely random, but again the low number of samples makes those results preliminary.

\begin{table}[h!]
\begin{center}
 \begin{tabular}{||c c c c||} 
 \hline
  & HSQC & HMBC & HSQC + HMBC \\ 
 \hline\hline
 Pure compounds & 82.56\% & 91.86\% & 90.84\% \\ 
 \hline
 Mixtures & 4/15 & 13/15 & 10/15 \\
 \hline
\end{tabular}
\caption{Accuracy of predicting compound classes for pure compounds, using cross-validation (first row) and predicting compound classes in mixtures (second row).}
\label{table:pure_and_mixtures}
\end{center}
\end{table}
 

\begingroup
\setlength{\tabcolsep}{3pt} 
\begin{table}[h]
\begin{center}
  \begin{tabular}{ || l | c | c | c || l | c | c | c || l | c | c | c || }
    \hline
     Indole & HSQC & HMBC & \pbox{20cm}{HSQC+\\HMBC} & Steroid & HSQC & HMBC & \pbox{20cm}{HSQC+\\HMBC} & \pbox{20cm}{Fatty\\acid} & HSQC & HMBC & \pbox{20cm}{HSQC+\\HMBC}  \\ \hline
    Pure & X & X & X & Pure & X & X & X & Pure & X & X & X \\ \hline
    +1 & F & X & X & +1 & F & X & X & +1 & X & X & X \\ \hline
    +2 & X & X & X & +2 & I & I & X & +2 & I & X & X \\ \hline
    +3 & F & X & F & +3 & F & I & X & +3 & I & X & S \\ \hline
    +4 & F & X & X & +4 & F & X & X & +4 & S & X & S \\ \hline
    +5 & X & S & S & +5 & X & X & X & +5 & S & X & S \\ \hline
  \end{tabular}
\end{center}  
\caption{Performance of finding substructures in mixtures. The columns headed "Indole," "Steroid," and "Fatty Acid" indicate which pure compound was artificially mixed. Each line represent an additional compound for the mixtures so that +1, +2, +3, +4 and +5 stands for the increasing complexity. An X in the columns means a correct detection, whereas I, F, S indicate that this mixture was wrongly detected as an indole, fatty acid, or steroid respectively. }
\label{table:performance_mixtures}
\end{table}
\endgroup

\section{Discussion}
\label{sec:discussion}

The results show that it is possible to determine substructures from 2D NMR spectra for pure compounds and also for mixtures as is shown tentatively shown using artificial data. In particular, the use of HMBC spectra, which has not been reported in the literature, gives better results for this purpose in comparison to HSQC spectra alone. By using an approach that enable no specific processing, this generates the advantage of a wide range of data to use in future work, including spectra coming from the literature and reported, e. g. PDFs in the supplemental information from peer-reviewed publications. This approach uses the spectral images as they are reported, without requiring any specific processing. SMART-NMR \cite{pmid29079836, doi:10.1021/jacs.9b13786}, on the other hand, uses a CNN with a similar architecture to the noted approach, but it aims at finding similarities between complete compounds. It does this by using HSQC spectra only and it requires an uniform processing pipeline for the spectral data.

Considering the mixture analysis evaluated here using a system of increasing complexity for each substructure case, we could prove some consistency (Table~\ref{table:performance_mixtures}). In all cases, we were to find the target substructure and more. Specifically, by including the data from bmse000544 (cholesteryl palmitate) as the fifth additional compound, we were able to detect its steroidal fragment in every case. The purine fragment in bmse000060 (adenine) and  bmse000061 (adenosine) might have shown to be similar to the correlation network of the indole target fragment, and so they were also identified in mixtures.

With respect to the use of HMBC data, it is noted that it improves results compared to HSQC, which is in line with the expectations. As a HSQC spectrum delivers correlations through $^1J_{CH}$, the information is limited to hydrogenated carbon positions. This is an improvement from the use of 1D $^1$H NMR data alone since the spectral window for peak dispersion is greatly increased and now it is possible to add $^{13}$C NMR data in comparison. This concept is well explored elsewhere \cite{pmid29079836,doi:10.1002/mrc.2486}. By including HMBC as a source of information, not only unique positions are considered but also are the correlation across different positions and so fragments of the whole structure of the detected compounds. Still, $^3J_{CH}$ data end up delivering information on the quaternary carbon atoms, invisible at HSQC data. Overall, it can be understood that the addition of HMBC into compound identification pipelines as an additional orthogonal dimension yields confidence into atom configuration. To note, the well-established COLMAR web-source (complex mixtures by magnetic resonance) used for NMR-base metabolomics also uses mainly HSQC data for matching a database for compound identification. An obvious limitation, though, is the lower sensitivity of the HMBC experiment compared to the HSQC experiment and the consequent instrument time it requires to obtain a rich spectrum. A possible solution would be the use of TOCSY-based spectra \cite{Bingol2016} rather than the HMBC, but this work has aimed to tackle the long rang correlation which could deliver bigger fragments without relying too much on $^3J_{CH}$ dihedral angles. TOCSY are one potential direction for future research. With the advent of nested NMR experiments  \cite{ClaridgeNOAH2019}, not to mention low volume and cryogenic probe technology, instrument time might be reduced due to increasing sensitivity gain.


\section{Conclusion}

In this work, it is demonstrated that it is possible to use a CNN to identify substructures in NMR spectra of mixtures. This is done using HSQC and HMBC spectra separately and in combination, which represents an improvement over previous work. It is shown that it is possible to reliably detect substructures in pure compounds, using a simple network and also tentatively, in mixtures. Furthermore, it is shown that this can work for mixtures when trained on pure compounds only. HMBC data and the combination of HMBC and HSQC show better results than HSQC alone. The use of an image-based approach enables the use of data from different sources. Further work will look at the use of HSQC-TOCSY spectra and overall, the use of more diverse types, and data, and variations on the initial architecture can be tried. The use of auto-encoders, subsampling architectures, like U-Net, or generative models are some of the possibilities.

\section*{Conflicts of interest}
There are no conflicts of interest in this work.

\section*{Supporting Information}

Data and code are available at \url{https://github.com/stefhk3/substructuresnmr}.

\printendnotes

\bibliography{main}



\end{document}